# Cross-sectional imaging of speed-of-sound distribution using photoacoustic reversal beacons

Yang Wang, Danni Wang, Liting Zhong, Yi Zhou, Qing Wang, Wufan Chen[*], Li Qi[*]

School of Biomedical Engineering, Southern Medical University, 1023 Shatai Rd., Baiyun District, Guangzhou, Guangdong 510515, China

Guangdong Provincial Key Laboratory of Medical Image Processing, Southern Medical University, 1023 Shatai Rd., Baiyun District, Guangzhou, Guangdong 510515, China

Guangdong Province Engineering Laboratory for Medical Imaging and Diagnostic Technology, Southern Medical University, 1023 Shatai Rd., Baiyun District, Guangzhou, Guangdong 510515, China

[*] Corresponding authors: W. Chen, chenwf@smu.edu.cn, L. Qi, qili@smu.edu.cn.

**Abstract:** Photoacoustic tomography (PAT) enables non-invasive cross-sectional imaging of biological tissues, but it fails to map the spatial variation of speed-of-sound (SOS) within tissues. While SOS is intimately linked to density and elastic modulus of tissues, the imaging of SOS distribution serves as a complementary imaging modality to PAT. Moreover, an accurate SOS map can be leveraged to correct for PAT image degradation arising from acoustic heterogeneities. Herein, we propose a novel approach for SOS reconstruction using only PAT imaging modality. Our method is based on photoacoustic reversal beacons (PRBs), which are small light-absorbing targets with strong photoacoustic contrast. We excite and scan a number of PRBs positioned at the periphery of the target, and the generated photoacoustic waves propagate through the target from various directions, thereby achieve spatial sampling of the internal SOS. We formulate a linear inverse model for pixel-wise SOS reconstruction and solve it with iterative optimization technique. We validate the feasibility of the proposed method through simulations, phantoms, and *ex vivo* biological tissue tests. Experimental results demonstrate that our approach can achieve accurate reconstruction of SOS distribution. Leveraging the obtained SOS map, we further demonstrate significantly enhanced PAT image reconstruction with acoustic correction.

## 1. Introduction

Photoacoustic tomography (PAT) offers a unique method for measuring optical absorption through ultrasonic detection. It achieves imaging of the internal tissue structures and functions by receiving and processing photoacoustic wave signals excited by light [1-5]. Due to its imaging characteristics in terms of molecular and functional contrast, as well as imaging depth, PAT has garnered increasing interest in the fields of pre-clinical and clinical imaging [6-8]. PAT employs computational image reconstruction, its image formation relies on estimating the spatial locations of the imaging objects based on the time of flight (TOF) of the photoacoustic waves [9-11]. Since TOF is determined by the speed-of-sound (SOS) distribution along the wave propagation path, tissue SOS heterogeneity will affect the accurate reconstruction of PAT images [12-14]. Therefore, SOS correction is a critical issue for improving PAT image quality.

In medical diagnosis, SOS of biological tissues is a useful physical parameter that reflects macroscopic density properties. Different tissue types possess distinct SOS values, e.g. the SOS of skin ranges from 1498 to 1540 m/s, muscles from 1500 to 1610 m/s, and bone from 1630 to 4170 m/s [15]. Therefore, the measurement of SOS can be utilized to differentiate between various tissues. Ultrasound tomography, or UST, is the only method to image the distribution of SOS within tissue noninvasively. For example, Li *et al.* reconstructed the SOS map of breast tissue in breast cancer patients using UST, enabling a clear separation between dense tumor lesions and adipose tissue [16]. On the other hand, the SOS of tissues and organs may also vary across different disease stages. As demonstrated by Boozari *et al.*, the detection of SOS values can quantify the degree of liver fibrosis, thereby aiding in assessing the prognosis and treatment of patients with chronic viral hepatitis [17].

Unlike UST, PAT imaging cannot measure tissue SOS distribution. Therefore, it loses the information of tissue SOS and the image reconstruction quality is heavily affected by SOS heterogeneity. A uniform SOS is often assumed for PAT image reconstruction, and autofocus methods have been introduced to iteratively determine an optimal uniform SOS by evaluating a cost function that describes image or signal quality [18-20]. These methods can only partially mitigate image distortions within a selected region of interest. Alternatively, SOS corrected PAT image reconstruction methods have been proposed. Jose *et al.* proposed a half-time reconstruction method, which artificially selects an empirical SOS value and discards time-domain data far from the transducer as distorted measurements [21]. This method is simple and fast due to data truncation, but it sacrifices a significant amount of measurement information. Cai *et al.* introduced a feature-coupled approach that simultaneously reconstructs the SOS map and PAT image by maximizing a similarity metric [22]. These joint reconstruction methods require good initial values and may easily get trapped in local optima. Lafci et al. manually segmented the image into foreground and background regions, and assigned different SOS values before image reconstruction [23]. This method improves image quality compared to fixed single-SOS approaches but introduces new challenges in image segmentation. Zhang et al. developed a dual-modality integrated device combining ultrasound imaging and PAT, which segments the sample and coupling medium and searches for the optimal SOS based on ultrasound pulse-echo signals to achieve adaptive SOS correction [24]. This method avoids the SOS uncertainty but relies on additional ultrasonic imaging equipment. Currently, there is no method that solely relies on PAT imaging to measure the SOS distribution [25-27].

In this paper, we propose a novel method for SOS imaging using only PAT imaging modality. This method spatially samples the SOS within the imaging object by placing and exciting tiny imaging targets with strong photoacoustic absorption outside the object, which we term photoacoustic reversal beacons (PRBs). When excited by laser pulses, the photoacoustic wave generated by a PRB propagates through the entire imaging object. Therefore, the total TOF of a PRB signal reaching a transducer element can be regarded as the integral of TOFs along its propagation path. By scanning and exciting the PRB around the imaging object, a linear SOS reconstruction model can be constructed based on the TOF information and propagation paths of the PRB signals. Subsequently, an inversion algorithm is employed to reconstruct the cross-sectional SOS at pixel level. To validate the effectiveness of the proposed method, we conducted numerical simulations, phantom experiments, and *ex vivo* biological tissue experiments using a ring-array PAT system. The experimental results demonstrate that our PRB method can robustly and accurately obtain the spatial distribution of SOS noninvasively. After obtaining the SOS of the imaging object, we further use this information to correct for tissue acoustic heterogeneity of PAT image reconstruction. Experimental PAT imaging results based on our SOS correction method exhibit significant image resolution improvement.

## 2. Methods

### A. Principle of photoacoustic reversal beacon

Our SOS reconstruction approach, as illustrated in Fig. 1(a), involves positioning a small photoacoustic absorbing target, which we referred to as the photoacoustic reversal beacon, or PRB, outside the imaging target. When the PRB and the imaging object are simultaneously excited by a laser, photoacoustic signals arise from both. The photoacoustic signal emanating from PRB traverses the imaging object before being captured by a transducer array. During its propagation, the wavefront of the PRB signal undergoes alterations due to the acoustic heterogeneity of the intervening tissues. Consequently, this transformed wavefront implicitly encodes the SOS information of the target, which is then leveraged for further analysis.

Obviously, relying solely on the signal from a single PRB is insufficient for reconstructing the SOS distribution. Therefore, as illustrated in Fig. 1(b), we achieve comprehensive coverage of the imaging object by positioning and exciting the PRB at various locations peripheral to the object. By acquiring the TOF information of the PRB at each location, we establish a linear mathematical model that relates SOS to the PRB's TOF to reach each transducer. Subsequently, we employ an inversion algorithm to solve for the pixel-wise SOS distribution within the scanned field of view. Finally, based on the obtained accurate SOS map, we can perform precise reconstruction of PAT images.

To separate the PRB signal from the target signal within the raw data, the PRB should possess highly absorbent optical properties and its size should be minimized to enhance the detectability of its signal. In this way, the PRB manifests as a short term, large amplitude signal in the acquired time-series data. Consequently, we can easily identify the PRB signal and calculate the TOF of PRB signal reaching each transducer element.

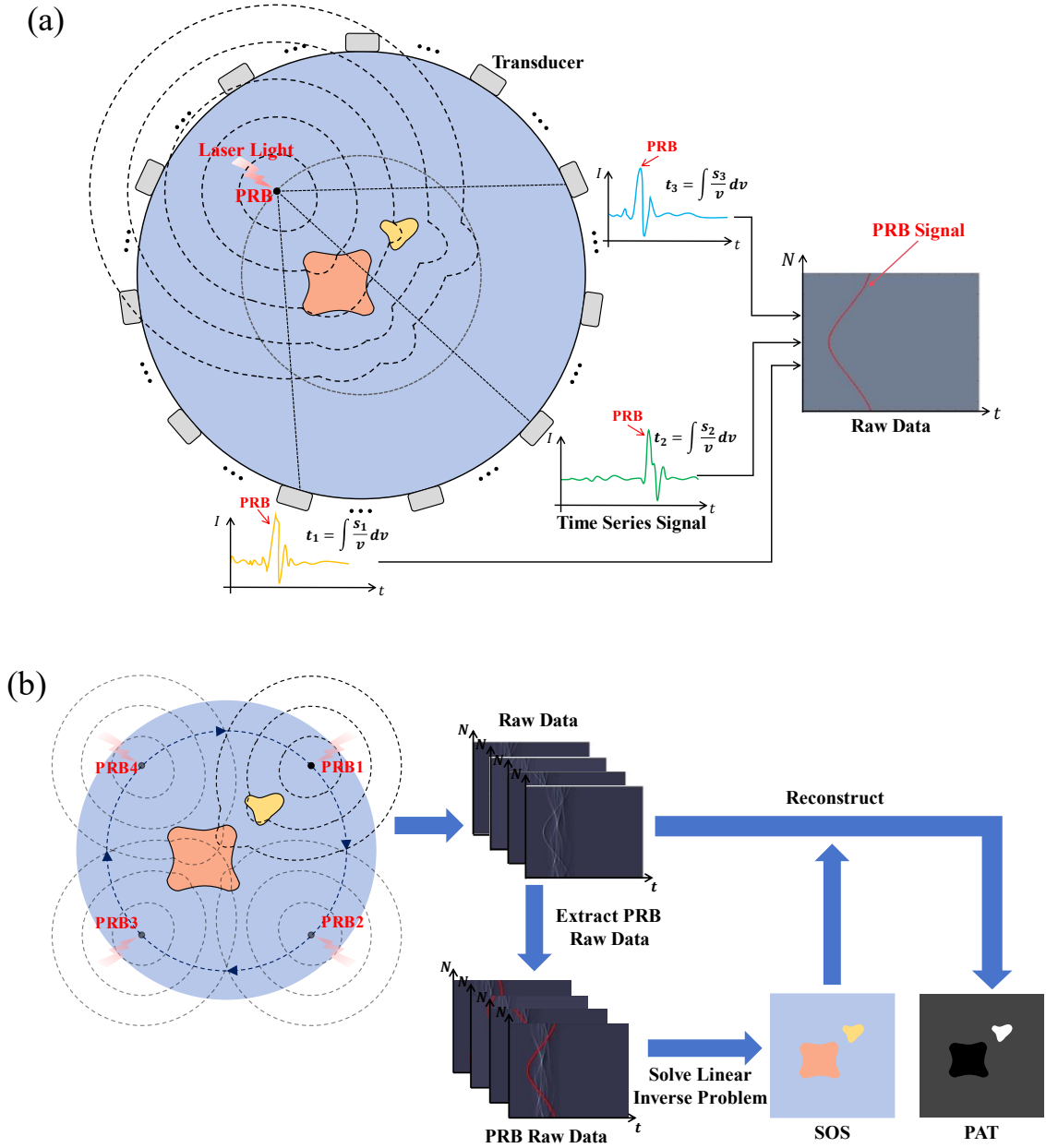

**Fig. 1.** Cross-sectional SOS imaging based on PRB. (a) The PRB signal and its acquisition process. Due to its strong light absorption, the signal of PRB appears as a short term, large amplitude signal. (b) Multi-position scanning of PRB realizes SOS imaging. The photoacoustic signal excited by PRB at different positions is fed into a linear inverse problem to reconstruction the SOS distribution. The obtained SOS map can be used to correct for acoustic heterogeneity during PAT image reconstruction.

### B. Linear inverse model for SOS reconstruction

Here we present the linear inverse model for obtaining a SOS distribution by using PRB scanning. To simplify the computational process, we neglect acoustic refraction, assuming that the ultrasonic waves propagate without deviation during their travel. We regard the TOF of photoacoustic waves as a line integral of the product of the propagation path operator and the reciprocal of the SOS, thus treating the inverse of SOS as a linear function of depth or distance. As a result, the TOF $\tau$ from a specific PRB position to a transducer element can be regarded as the linear integral of the TOF at each spatial position along the propagation path $S$,

$$\tau = \int_S \frac{ds}{v(s)}, \tag{1}$$

where $v(s)$ is the velocity of the photoacoustic wave at position $s$.

Herein, we discretize the propagation path into pixels, allowing the TOF to be treated as a linear integral of the flight times within each pixel along the path. The TOF within a given pixel is, in turn, related to the propagation distance within that pixel and the local speed of sound at that pixel. This approach forms the basis of our mathematical model, as illustrated in Fig. 2:

$$y = \mathbf{W}x, \tag{2}$$

where, $y$ is the recorded TOF from each PRB to all transducer elements, $x$ is the reciprocal of the SOS distribution at each location in space, and $\mathbf{W}$ is the propagation path operator of the PRB with a dimension of ($M *N$, K), where $K$ is the discretized spatial locations. As shown in fig. 2, each element in $\mathbf{W}$ represents the signal traverse distance within the pixel indexed as $k$ on the propagation path when the signal emitted from the $n$-th PRB propagates to the $m$-th transducer.

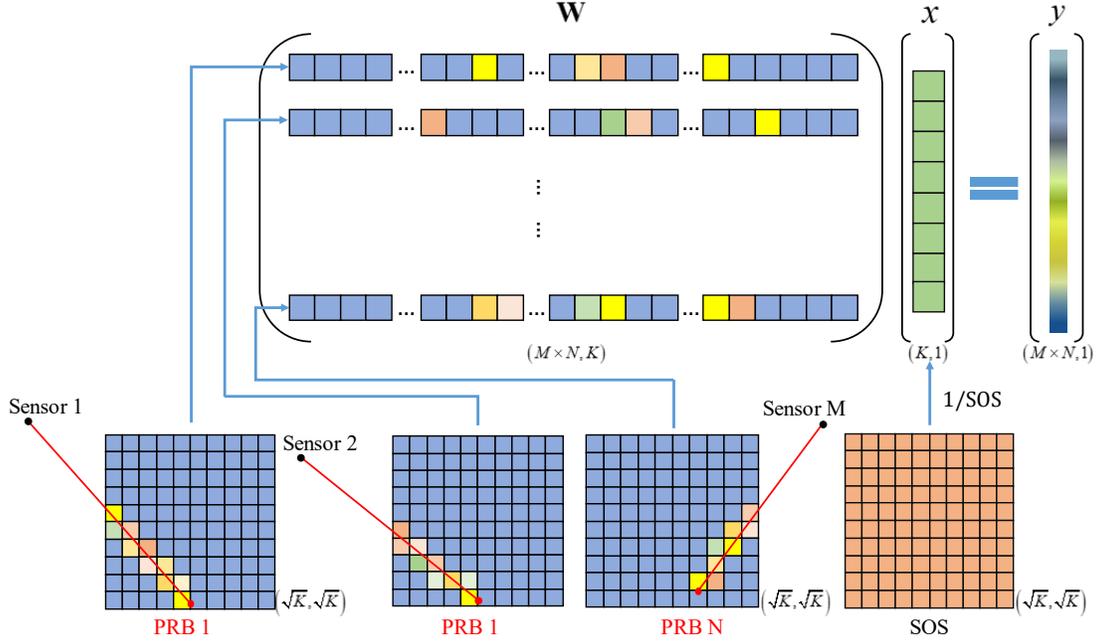

**Fig. 2.** Linear forward model of PRB-based SOS imaging. *x*, SOS map. **W**, propagation path operator. *y*, TOF of PRB. Different colors in **W** represent the pass-through distance within the current pixel, with blue-purple denote zero. *M*, number of transducers. *N*, number of PRB positions. *K*, number of pixels.

### C. TOF calculation for PRB

The SOS reconstruction necessitates the calculation of TOF from each RPB to all transducer elements, as illustrated in Fig.3. For each PRB scan, we employ a Dynamic Programming (DP) algorithm based on graph theory [28] to extract the signal of PRB from the acquired raw data. This algorithm effectively compensates for signal fluctuations and acquisition errors that may arise during PRB scanning operation, thereby ensuring signal recognition continuity and accuracy. Then, the TOF of the photoacoustic waves excited at the current PRB position and reaching each transducer is calculated based on the sampling frequency of the data acquisition system and the extracted PRB signals. This results in the construction of a TOF vector $t = [\tau_1, \tau_2, \ldots, \tau_M]$, where *M* is the total number of transducer elements. Subsequently, by scanning the PRB in *N* positions around the imaging object, we are able to acquire *N* TOF vectors. These vectors are then concatenated into a TOF vector *y* with a dimension of (*M* * *N*, 1) to be used for solving the linear problem in Eq. (2).

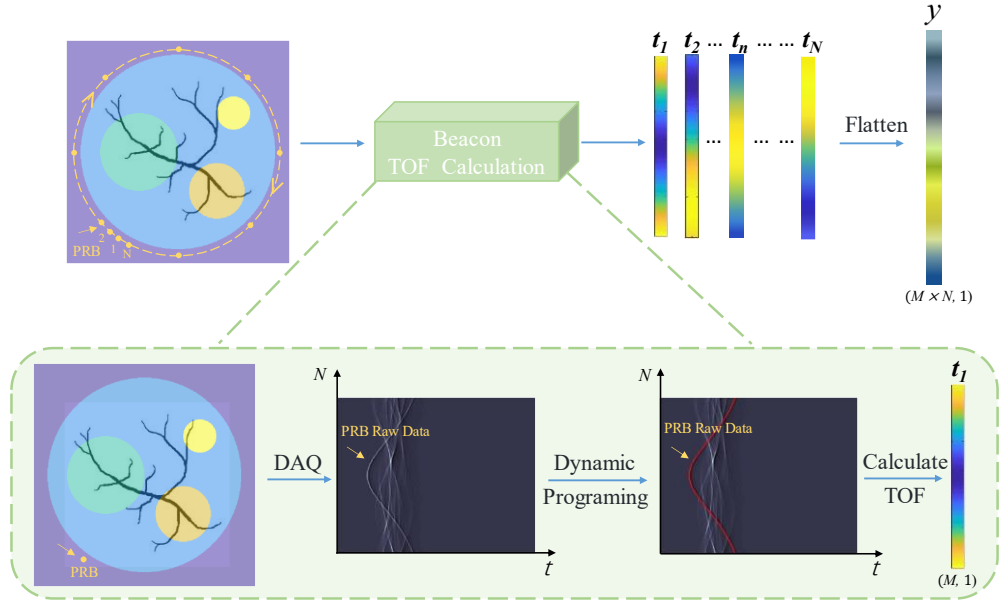

**Fig. 3.** PRB signal extraction and TOF calculation. The PRB trace identified in the original signal is used to calculate the TOF $t_n$ ($n$ = 1,2,3 … $N$) from the current PRB position to each detector. By obtaining $t_1$ to $t_N$ at $N$ positions around the object, it is vectorized for subsequent calculations.

To improve accuracy, we acquire the TOF of PRB propagating through only the background medium with no imaging object presented, and denoted this background TOF as $t_b$. We then use the background subtracted TOF $\Delta t = t_b - t$, to serve as the vector $y$, for solving $x$. As indicated in Fig. 4, by subtracting the background TOF information, we are able to effectively isolate the signal perturbation caused by the SOS of the imaging object.

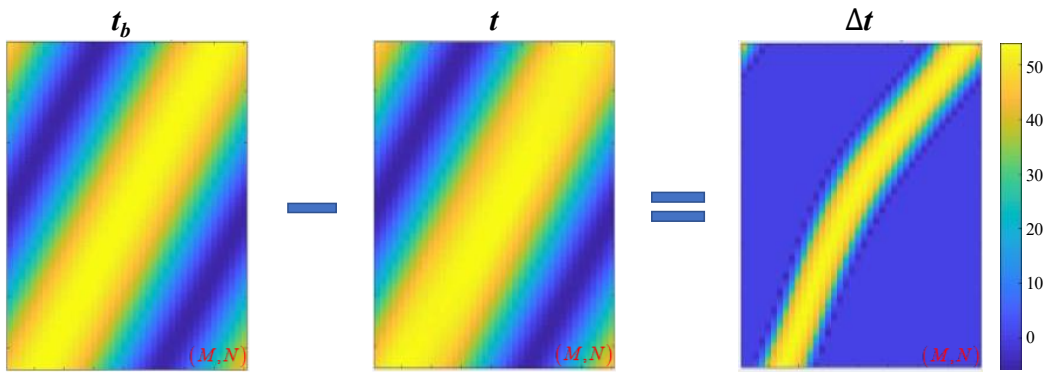

**Fig. 4.** Visualization of the SOS perturbation induced by the imaging object. $t$: TOF of the PRB with imaging object presented, $t_b$: TOF of the PRB of the background medium, $M$: the total number of scanned PRBs, $N$: the total number of transducer elements.

### D. SOS distribution reconstruction

For the linear problem in Eq. (2), given the recorded TOF vector $y$ and the propagation path operator **W**, an inversion algorithm can be employed to find a solution of the SOS distribution $x$. To avoid overfitting, regularization constraints can be incorporated into the solution process, as formulated below:

$$x^* = \arg\min\left\{\frac{1}{2}\|\mathbf{W}x - y\|^2 + \lambda\|x\|_2^2\right\}, \quad (3)$$

We adopt a $L_2$-norm regularization approach to prevent overfitting or excessively large parameter values, and $\lambda$ serves as the regularization parameter to balance the weight between the data fidelity term and the regularization term. To solve the above problem, we utilize the Nesterov's accelerated gradient descent algorithm with momentum [29] and setting the maximum iteration to 200.

### E. SOS corrected PAT image reconstruction with TI-MDAS

Delay and Sum (DAS) is a commonly used method for PAT image reconstruction [30,31]. This technique involves superimposing and reconstructing the received photoacoustic signals to recover the initial pressure image of the tissue. The basic principle of DAS can be expressed as follows:

$$P_{DAS}(k) = \sum_{k=1}^{K}\sum_{m=1}^{M} p(m, \tau(k)), \quad (4)$$

where, $P$ is the reconstructed PAT image, $p$ is the raw data acquired by the transducer, and $\tau(k)$ is the TOF at which the photoacoustic signal of pixel $k$ is captured. The traditional DAS algorithm relies on the assumption of uniform acoustic properties in biological tissues, hence adopting a constant SOS value during calculation. Therefore, SOS error accumulates during image reconstruction, ultimately manifesting as image distortions and artifacts, thereby compromising imaging quality [12-14, 32].

Based on the proposed PRB method, we can obtain a precise distribution of the internal SOS, which can be used to perform SOS-corrected PAT image reconstruction. To modify the traditional DAS algorithm, we introduce the TOF Interpolation-based Multi-SOS Delay-and-Sum (TI-MDAS) algorithm. Our TI-MDAS algorithm transcends the constraint of fixed SOS parameters by using:

$$\tau(k) = \sum_{k=1}^{K}\frac{s_k}{v_k}, \quad (5)$$

where, $s_k$ and $v_k$ represent the flight distance of the signal at the $k$-th target pixel and the velocity of reconstructing the photoacoustic signal at the same pixel, respectively. Compared to traditional DAS algorithms, the above operation enables precise SOS correction at the pixel level but requires a higher time cost in calculating the TOF. To address this issue, we first calculate the coarse-grid TOF under low-resolution conditions. Subsequently, we utilize a bicubic interpolation algorithm to interpolate the initial TOF onto a finer grid, thus significantly enhancing the reconstruction speed.

## 3. Experimental setup

### A. Cross-sectional PAT imaging system

In this work, we utilized a commercially available photoacoustic tomography system, the MSOT inVision128 system by iThera Medical GmbH, Germany. The detection component of the system comprises a circular transducer array consisting of 128 elements. This array spans a solid angle of 270 degrees, with a radius of 40.5 mm, and the active region features a width of 20 mm (Fig. 5(a)). The transducer has a central frequency of 5 MHz. The system operates at a sampling frequency of 40 MHz, and illumination is provided by an optical parametric oscillator (OPO) laser tunable from 680 nm to 980 nm.

### B. PRB and its scanning system

We employed a thin, black nylon wire with a diameter of 0.2 mm as the PRB. To achieve precise positioning of the PRB, we designed a mechanical PRB scanning device, which comprises a control unit and an execution unit (Fig. 5(b)). The control unit utilizes an Arduino microcontroller and the execution unit is a transmission system to enable rotational scanning of the PRB. The rotational movement provided by a stepper motor is transmitted via a timing belt and shaft to the bottom timing pulley, where it is fixed to a 3D-printed fixture that secures the nylon wire at both ends. The entire structure is mounted on the imaging frame of the MSOT system (Fig. 5(c)), which realizes precise positioning of the PRB. The scanning process of this work can be found in the Supplementary Materials

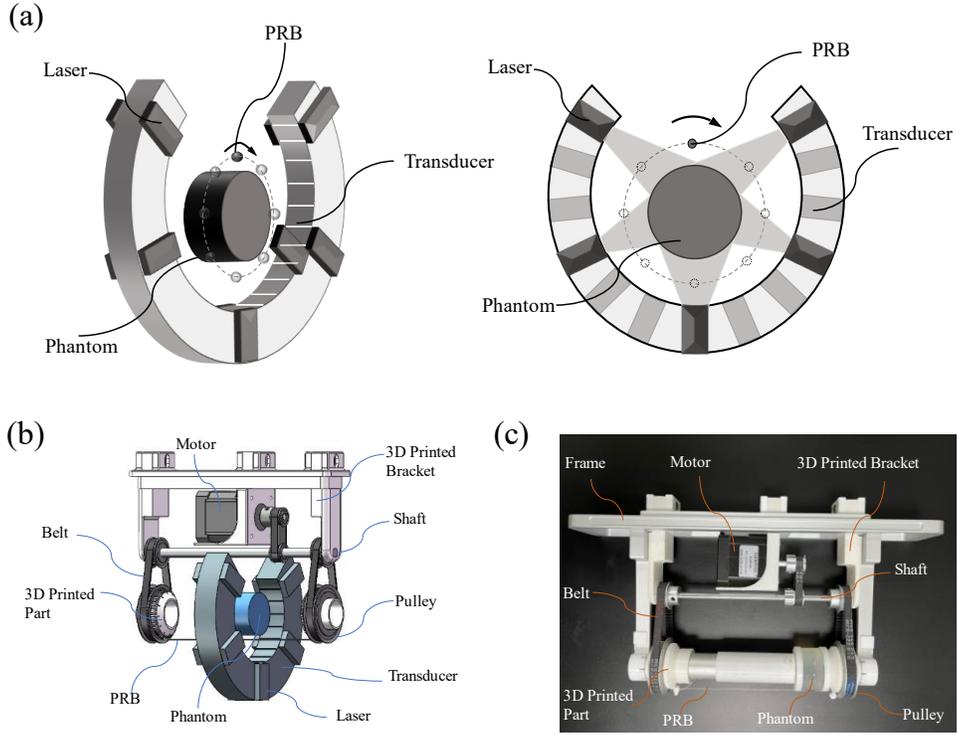

**Fig. 5.** PRB and its scanning system. (a) The imaging geometry of the used system with PRB. (b) Rendering of the PRB rotational scanning system. (c) Photograph of the assembled PRB rotational scanning device.

### C. Simulation experiment

To validate the effectiveness of the PRB method, we conducted simulation experiments utilizing the k-Wave toolbox [33]. To closely approximate real-world conditions, we adapted the numerical simulation parameters to match those of the MSOT system. The PRB was configured to rotate with a radius of 19 mm. The SOS of the surrounding medium was set to 1500 m/s, and the imaging field of view was defined as 40 mm$^2$.

We first introduced a single circular imaging target with a uniform SOS (Fig. 6(a)), setting its SOS to 1650 m/s and adjusting its diameter from 16 mm to 4 mm to evaluate SOS reconstruction based on our PRB method. Additionally, we designed a target containing four distinct SOS regions (Fig. 6(b)), and employed a vascular-mimicking structure as the absorbing structure for PAT imaging (Fig. 6(c)). This setup was used to verify the PAT image reconstruction performance after SOS correction.

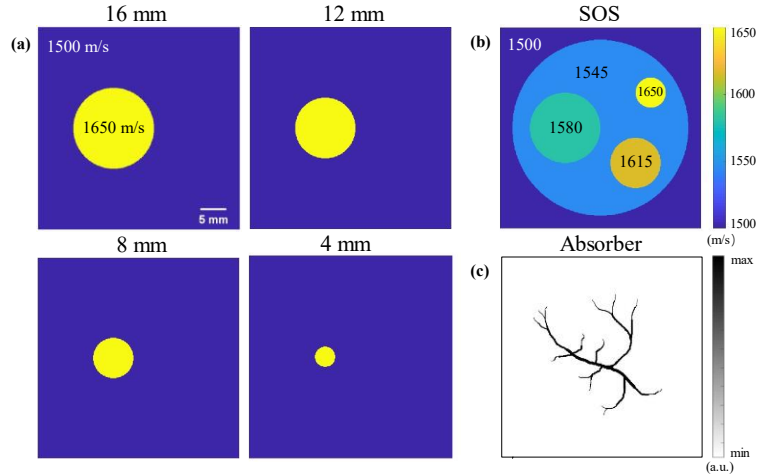

**Fig. 6.** Simulation experiment. (a) single target images with different target diameters. (b) multi-target image with various target diameters and SOSs. (c) Photoacoustic absorbing structure corresponds to (b).

### D. Phantom experiment

Tissue-mimicking phantom experiments were conducted on the used MSOT system. We fabricated acoustically heterogeneous phantoms with two different SOS media (Fig. 7(a)), where the outer medium was composed of 2.5% w/w agar mixed with water, yielding an SOS of approximately 1520 m/s (measured from the average echo time using a linear array ultrasound transducer). The inner medium was a mixture of 1.25% w/w agar and 70% w/w glycerol, with a measured SOS of approximately 1750 m/s. We obtained 4 phantoms with the diameter of the inner medium changed from 4 mm to 16 mm. Additionally, to verify the PAT imaging performance with SOS correction, a phantom was fabricated to uniformly embed 250 μm microspheres as targets for PAT imaging (Fig. 7(b)). Throughout the experiments, the temperature of the water surrounding the phantom was set to 26°C, thereby we set the SOS of water to 1500 m/s [34-37].

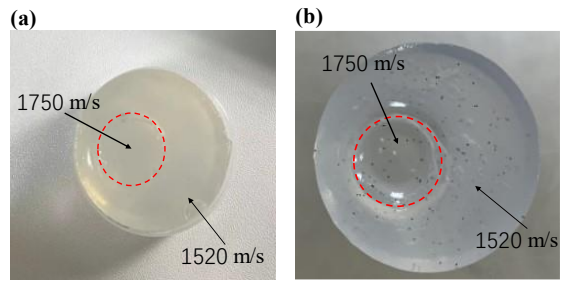

**Fig. 7.** (a) Photograph of a phantom with varying SOS distributions. (b) Photograph of the acoustic heterogeneous phantom embedded with black microspheres.

### E. Ex vivo biological tissue imaging experiment

To further validate the effectiveness of the PRB method, we conducted experiments on ex vivo biological tissues. Lamb and pork tissues were selected as the evaluation subjects due to their multilayered structure of fat and muscle. According to [15], the SOS of fat ranges approximately from 1450 to 1520 m/s, while that of muscle ranges from 1500 to 1610 m/s. Therefore, the notable difference in SOS between fat and muscle tissues makes them ideal biological samples for demonstrating acoustic heterogeneity. Additionally, to verify the imaging performance of PAT after SOS correction, three carbon rods with a diameter of 500 μm were inserted into an ex vivo tissue sample, serving as the targets for PAT imaging.

### F. Quantitative metrics

To evaluate the reconstruction results of SOS distribution using the PRB method, we have selected performance metrics [38,39], including Peak Signal-to-Noise Ratio (PSNR) and Structural Similarity (SSIM), which are defined as follows:

$$PSNR = 10 \cdot \log_{10}\left(\frac{MAX^2}{MSE}\right) \quad (6)$$

$$SSIM(x,y) = \frac{(2\mu_x\mu_y + C_1)(2\sigma_{xy} + C_2)}{(\mu_x^2 + \mu_y^2 + C_1)(\sigma_x^2 + \sigma_y^2 + C_2)} \quad (7)$$

where, MAX represents the maximum possible range of pixel values, while MSE is the average of the squared differences between corresponding pixels in two images. $\mu_x$ and $\mu_y$ is the mean of images $x$ and $y$, $\sigma_x^2$ and $\sigma_y^2$ are their variances, and $\sigma_{xy}$ are their covariance, respectively.

## 4. Results

### A. Simulation experiment results

In the simulation experiment of single target image, while adjusting the diameter of the target, we also investigated the impact of the number of scanned PRBs, i.e. $N$, on the reconstruction of SOS distribution and $N$ was selected as 10, 20, 40, and 80 (more results are given in Supplementary Materials). The results of SOS reconstruction under these conditions are shown in Fig. 8. The results reveal that the SOS target can already be identified when $N$ is 10, even though there are also notable artefacts. However, as the number of PRB increases, the artefacts are significantly suppressed, and the SOS reconstruction effect is improved. When $N = 80$, the artefacts nearly disappear, and targets of varying sizes, ranging from 16 mm down to 4 mm, can all conform to their identical size. We also calculated the SOS reconstruction accuracy within the target and background regions, and the results are shown in Table 1 and Table 2 respectively. As can be seen, as the number of scanned PRB positions increases, the error in regional SOS reconstruction gradually decreases. By comparing the error of different target sizes, it can be seen that smaller reconstruction errors have been achieved for larger-sized targets.

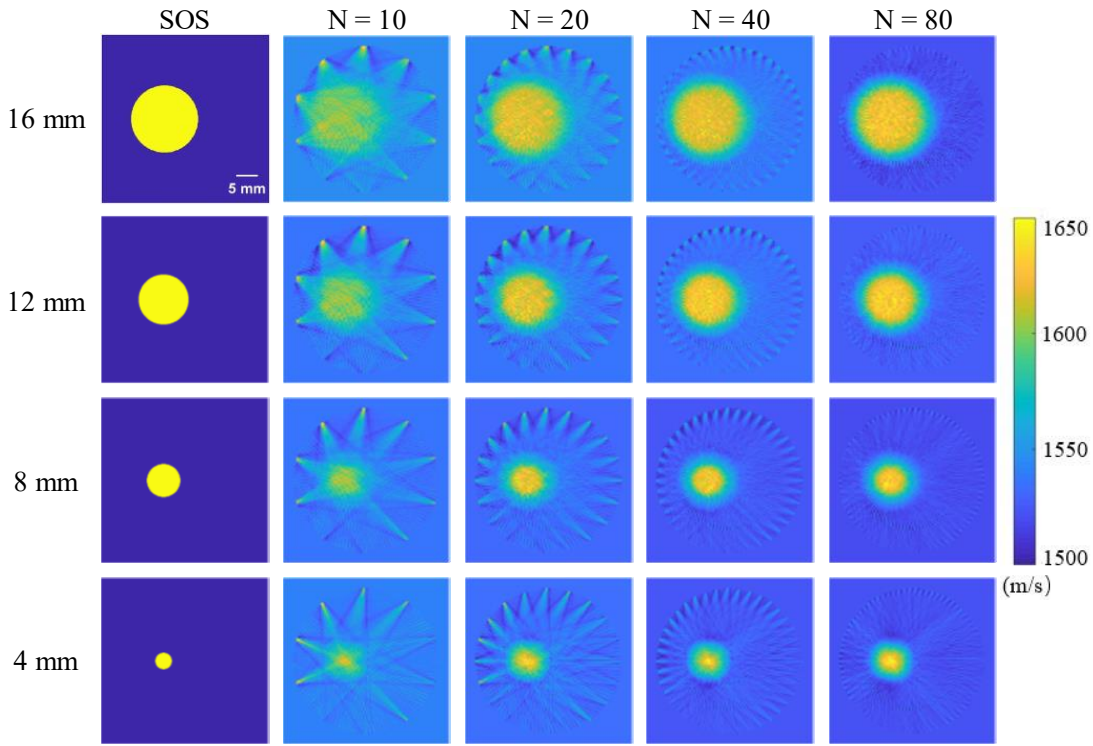

**Fig. 8.** Simulation experiment. Reconstruction results of SOS distribution for single target image with different target sizes under different numbers of scanned PRB positions.

**Table 1**
SOS reconstruction accuracy of the target region of the single target image in the simulation experiment. The ideal SOS of the region is 1650 m/s. $N$, the number of scanned PRB.

| N | Target diameter = 16 mm | | 12 mm | | 8 mm | | 4 mm | |
|---|---|---|---|---|---|---|---|---|
| | Mean ± Std. | % Err. | Mean ± Std. | % Err. | Mean ± Std. | % Err. | Mean ± Std. | % Err. |
| 10 | 1593.02±19.01 | 3.45% | 1581.42±17.08 | 4.16% | 1566.16±15.03 | 5.08% | 1527.23±5.14 | 7.44% |

| 20 | 1609.03±17.84 | 2.48% | 1599.32±16.22 | 3.07% | 1584.16±16.81 | 3.99% | 1533.53±5.07 | 7.06% |
| 40 | 1618.34±16.58 | 1.92% | 1610.79±15.27 | 2.38% | 1597.46±18.31 | 3.18% | 1540.32±7.12 | 6.65% |
| 80 | 1619.11±16.43 | 1.87% | 1612.29±15.31 | 2.29% | 1598.28±17.33 | 3.13% | 1540.07±4.03 | 6.66% |

**Table 2**

SOS reconstruction accuracy of the background region of the single target image in the simulation experiment. The ideal SOS of the background region is 1500 m/s. *N*, the number of scanned PRB.

| N | Target diameter = 16 mm | | 12 mm | | 8 mm | | 4 mm | |
|---|---|---|---|---|---|---|---|---|
| | Mean ± Std. | % Err. | Mean ± Std. | % Err. | Mean ± Std. | % Err. | Mean ± Std. | % Err. |
| 10 | 1515.42±31.77 | 1.03% | 1509.68±24.25 | 0.65% | 1505.07±16.18 | 0.34% | 1501.54±6.05 | 0.10% |
| 20 | 1513.49±28.42 | 0.90% | 1508.64±22.38 | 0.58% | 1504.64±15.38 | 0.31% | 1501.53±6.26 | 0.10% |
| 40 | 1512.08±25.46 | 0.81% | 1507.75±20.37 | 0.52% | 1504.15±13.97 | 0.28% | 1501.51±6.31 | 0.10% |
| 80 | 1512.02±24.28 | 0.80% | 1507.65±19.19 | 0.51% | 1504.14±13.12 | 0.28% | 1501.50±5.88 | 0.10% |

In the simulation experiment of multi-target image, we also investigated the impact of the number of scanned PRB positions on the reconstruction of SOS distribution. The results are shown in Fig. 9(a). It can be observed that our PRB method exhibits excellent capability in reconstructing SOS under complex conditions. Specifically, when N=20, the target areas begin to be identifiable. Additionally, through the SOS profile analysis shown in Fig. 9(b), it can be seen that the obtained SOS distribution is close to the ground truth. Moreover, at the edge of the SOS profile, which is more sensitive to noise as indicated by the red arrow in Fig. 10(b), the SOS results become increasingly accurate as N increases. We calculated the SOS reconstruction error within the multi-speed targets, and the results are shown in Table 3. Similarly, as the number of scanned PRB positions increases, the errors in regional SOS distribution all show a decreasing trend, and the maximum percentage error is within 3.59% even at N = 10. Furthermore, by comparing the reconstruction accuracy between single-target and multi-target images, it can be observed that the reconstruction error appears to be smaller in complex sound speed environments. This is attributed to the fact that under the single-target condition, as the diameter of the medium decreases, the proportion of sound speed error in the cost function decays exponentially, while the background error becomes the dominant factor in the iterative optimization process. In contrast, under the multi-target setting, the overall sound speed configuration tends to be more uniform, mitigating the prominence of background effects. Consequently, the overall error distribution is more stable, which also reflects the potential of our algorithm in realistic and complex sound speed environments.

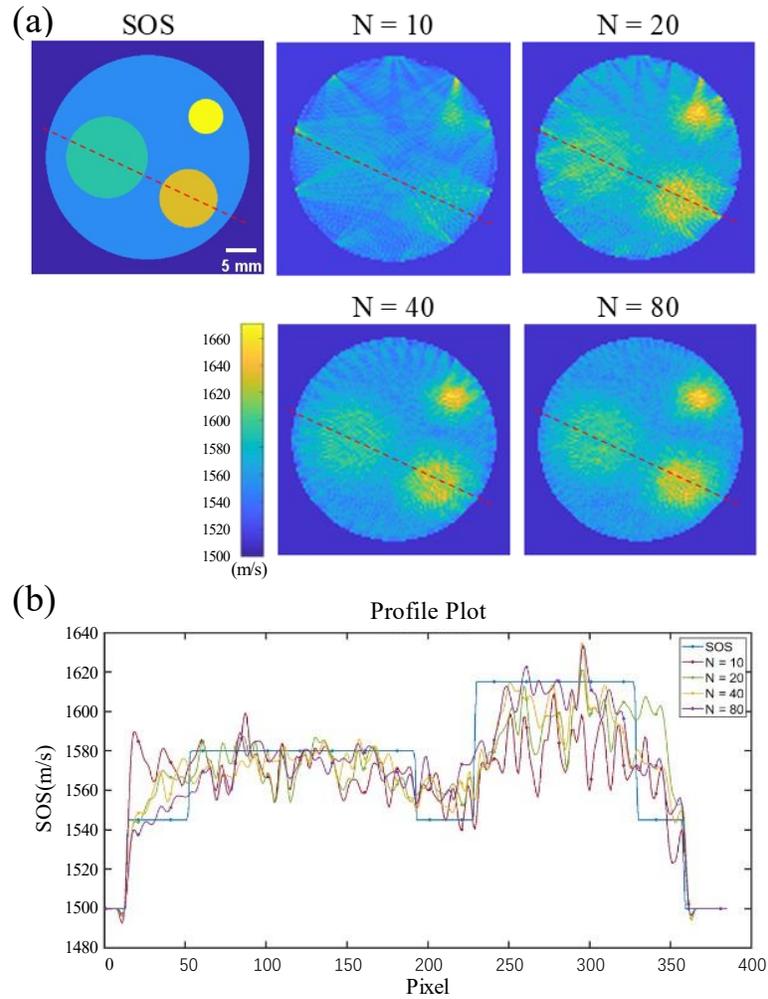

**Fig. 9.** Simulation experiment. (a) SOS reconstruction results for multi-target image under different numbers of scanned PRB. (b) The SOS profiles along the image indicated by the red dashed line in (a).

**Table 3**

SOS reconstruction accuracy of the multi target image in the simulation experiment. *N*, the number of scanned PRB.

| N | 1545 m/s region | | 1580 m/s region | | 1615 m/s region | | 1650 m/s region | |
|---|---|---|---|---|---|---|---|---|
| | Mean ± Std. | % Err. | Mean ± Std. | % Err. | Mean ± Std. | % Err. | Mean ± Std. | % Err. |
| 10 | 1552.11±17.53 | 0.46% | 1563.43±10.92 | 1.05% | 1580.16±15.57 | 2.16% | 1590.82±17.62 | 3.59% |
| 20 | 1552.01±13.98 | 0.45% | 1569.45±9.82 | 0.67% | 1588.94±14.48 | 1.61% | 1596.59±15.97 | 3.24% |
| 40 | 1551.66±12.34 | 0.43% | 1571.75±9.41 | 0.52% | 1594.09±15.18 | 1.29% | 1603.51±17.98 | 2.82% |
| 80 | 1551.67±11.64 | 0.43% | 1572.27±8.83 | 0.49% | 1594.03±14.59 | 1.30% | 1601.53±16.81 | 2.94% |

Furthermore, we present the quantitative analysis data of SSIM and PSNR for both single-target and multi-target images in Table 4 and Table 5 respectively. As can be seen from the tables, as the number of scanned PRB positions increases, the accuracy of SOS reconstruction gradually improves. Specifically, SSIM exhibits a positive correlation with increasing target size, whereas PSNR demonstrates an inverse relationship. This disparity stems from their distinct emphases: SSIM primarily assesses overall visual similarity, and as the target shrinks, the background proportion correspondingly swells, enhancing SSIM's ability to gauge the method's efficacy in reconstructing the comprehensive sound velocity profile of the target. Conversely, PSNR shines in evaluating the method's precision within specific, internal sound velocity regions. Notably, in the multi-target results, both PSNR and SSIM metrics affirm the exceptional performance of the PRB method in these complex scenarios, underlining its prowess in addressing real-world challenges.

**Table 4**

SSIM of SOS reconstruction for single-target and multi-target images under different numbers of scanned PRB.

| N | 16 mm | 12 mm | 8 mm | 4 mm | Multi-target |
|---|---|---|---|---|---|
| 10 | 0.69 | 0.76 | 0.83 | 0.93 | 0.73 |
| 20 | 0.72 | 0.77 | 0.84 | 0.94 | 0.77 |
| 40 | 0.78 | 0.82 | 0.87 | 0.95 | 0.81 |
| 80 | 0.83 | 0.87 | 0.90 | 0.96 | 0.83 |

**Table 5**

PSNR of SOS reconstruction for single-target and multi-target images under different numbers of scanned PRB.

| N | 16 mm | 12 mm | 8 mm | 4 mm | Multi-target |
|---|---|---|---|---|---|
| 10 | 34.36 | 33.61 | 31.57 | 30.25 | 39.00 |
| 20 | 36.00 | 34.85 | 31.94 | 30.31 | 40.69 |
| 40 | 37.23 | 35.82 | 32.19 | 30.36 | 41.70 |
| 80 | 37.48 | 36.11 | 32.30 | 30.37 | 41.88 |

We performed PAT image reconstruction on the multi-target image using the proposed TI-MDAS algorithm based on the obtained SOS reconstruction result, the results are as shown in Fig. 10(a). In the DAS reconstruction results using a single SOS, we set the uniform SOS to 1540 m/s to achieve the best visual effect for the main structure of the imaged object. As comparison, our TI-MDAS algorithm corrects the structural splitting caused by SOS heterogeneity and obtains better image quality. Moreover, as can be seem from the image profile shown in Fig. 10(b), our TI-MDAS also achieved a better focusing effect on the main part of the blood vessel. Additionally, although the visual differences in the reconstructed PAT images are not significant as the number of scanned PRB increases, SSIM and PSNR analysis in Table 6 reveals that the quality of PAT images is still being optimized.

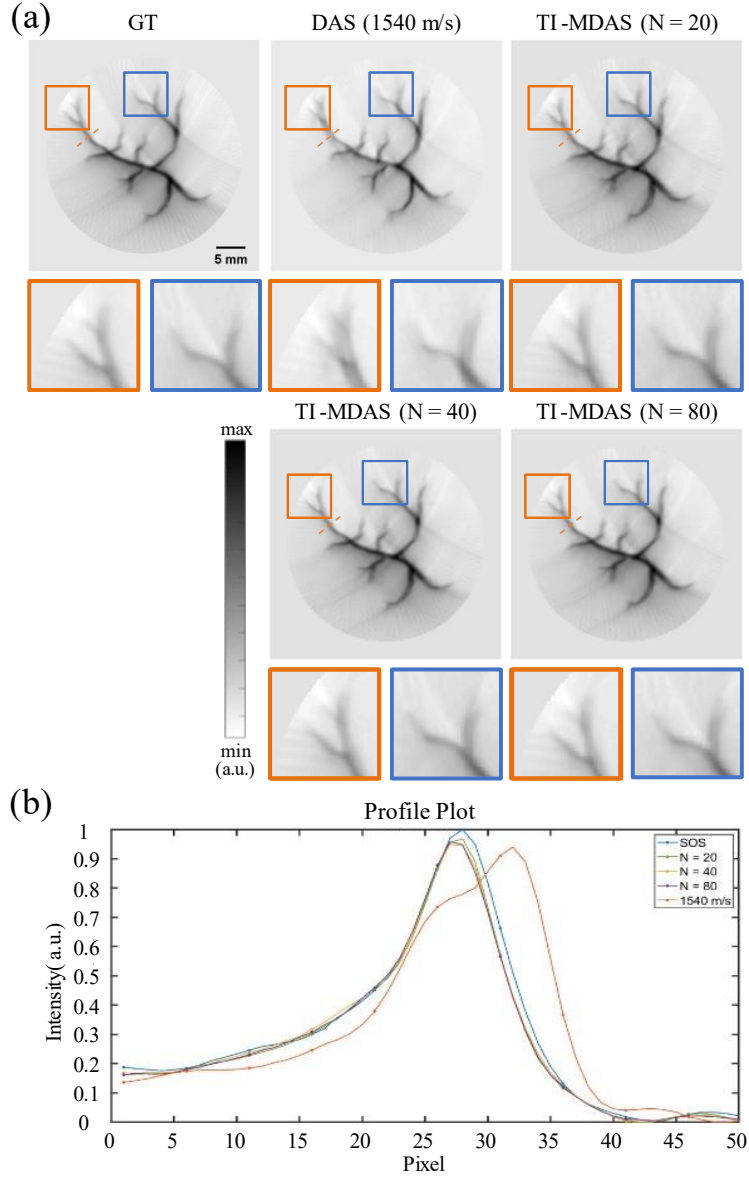

**Fig. 10.** PAT image reconstruction results using DAS algorithm (1540 m/s) and our TI-MDAS algorithm guided by SOS maps reconstructed by different numbers of scanned PRB. (b) The intensity profiles along the red dashed line in (a).

**Table 6**
SSIM and PSNR for PAT reconstruction using DAS algorithm (1540 m/s) and TI-MDAS algorithm guided by SOS maps reconstructed by different numbers of scanned PRB.

|      | 1540 m/s | N = 20 | N = 40 | N = 80 |
|------|----------|--------|--------|--------|
| SSIM | 0.9014   | 0.9592 | 0.9617 | 0.9619 |
| PSNR | 27.36    | 36.96  | 37.81  | 37.99  |

### B. Phantom experiment results

Fig. 11 shows the reconstructed results of the SOS distribution of the imaging phantoms. As can be seen, the position and size of the inserted regions with higher SOS have a very good correspondence with the actual image of the phantom shown in Fig. 7. Furthermore, it is evident that artifacts are predominantly distributed radially at the edge of the images. Notably, as the number of scanned PRB increases, these artifacts are significantly suppressed. Importantly, this trend remains consistent regardless of the varying sizes of targets with heterogeneous sound velocities, underscoring the robust potential of our method for accurate SOS imaging. As shown in Table 7 and Table 8, we also calculated the SOS reconstruction accuracy of the target and background in the phantom experiment. The reconstruction error of the phantom's SOS shows the same pattern as the simulation experiments, that is, as the number of beacon positions increases, the error continuously decreases. Due to the complementary area of the inner and outer regions, as the size of the internal sound velocity region changes, the error trends of the two also tend to complement each other.

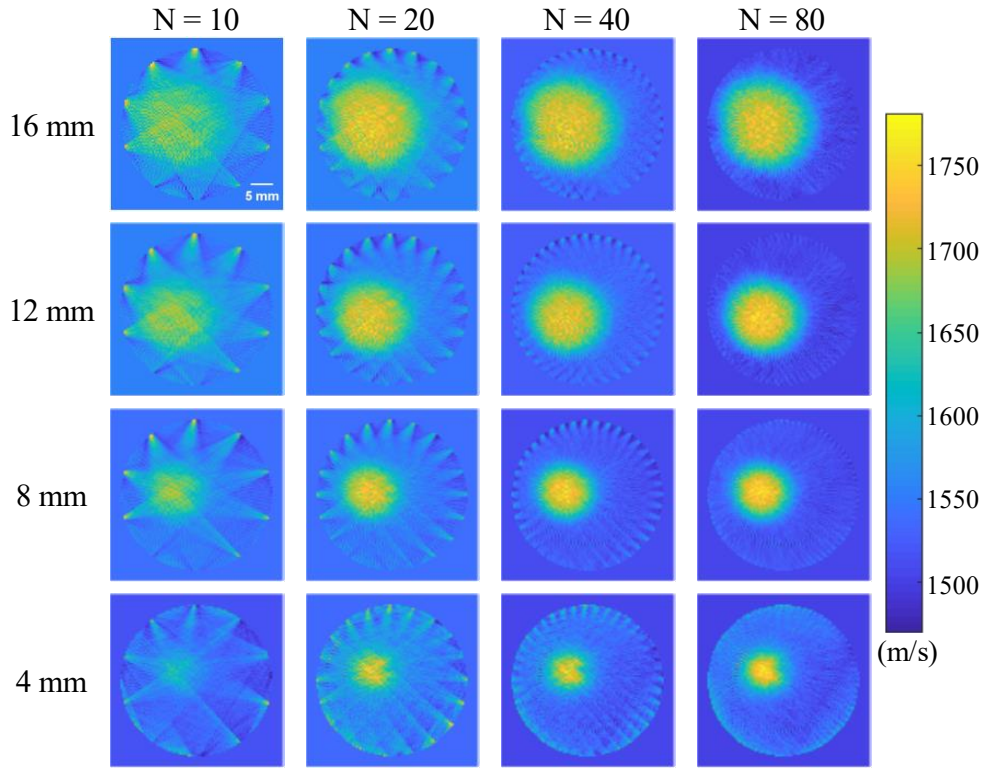

**Fig. 11.** Phantom experiment. Reconstruction results of SOS distribution for single target phantoms of different target sizes under different numbers of scanned PRB positions.

**Table 7**

SOS reconstruction accuracy of the target region of the single target image in the phantom experiment. The ideal SOS of the region is 1750 m/s. *N*, the number of scanned PRB.

| N | Target diameter = 16 mm | | 12 mm | | 8 mm | | 4 mm | |
|---|---|---|---|---|---|---|---|---|
| | Mean ± Std. | % Err. | Mean ± Std. | % Err. | Mean ± Std. | % Err. | Mean ± Std. | % Err. |
| 10 | 1689.70±40.55 | 3.45% | 1657.46±32.71 | 5.29% | 1627.87±24.56 | 6.98% | 1545.01±8.57 | 11.71% |
| 20 | 1714.25±37.20 | 2.04% | 1687.58±31.43 | 3.57% | 1656.07±24.35 | 5.37% | 1555.59±8.64 | 11.11% |
| 40 | 1730.78±33.56 | 1.10% | 1706.57±28.32 | 2.48% | 1676.76±22.32 | 4.19% | 1563.20±8.61 | 10.67% |
| 80 | 1732.48±31.66 | 1.00% | 1709.96±25.96 | 2.29% | 1680.34±18.76 | 3.98% | 1564.50±5.30 | 10.60% |

**Table 8**

SOS reconstruction accuracy of the background region of the single target image in the phantom experiment. The ideal SOS of the background region is 1520 m/s. *N*, the number of scanned PRB.

| N | Target diameter = 16 mm | | 12 mm | | 8 mm | | 4 mm | |
|---|---|---|---|---|---|---|---|---|
| | Mean ± Std. | % Err. | Mean ± Std. | % Err. | Mean ± Std. | % Err. | Mean ± Std. | % Err. |
| 10 | 1556.93±68.02 | 2.43% | 1535.45±51.66 | 1.02% | 1523.54±35.19 | 0.23% | 1512.62±12.67 | 0.49% |
| 20 | 1554.21±64.17 | 2.25% | 1534.80±50.74 | 0.97% | 1524.10±35.37 | 0.27% | 1513.15±12.02 | 0.45% |
| 40 | 1552.63±60.99 | 2.15% | 1533.89±49.72 | 0.91% | 1523.94±34.97 | 0.26% | 1513.46±11.30 | 0.43% |
| 80 | 1552.68±59.45 | 2.15% | 1533.62±48.47 | 0.90% | 1523.86±34.00 | 0.25% | 1513.38±10.68 | 0.44% |

The PAT image reconstruction results of the phantom with microspheres as the imaging object are shown in Fig. 12(a). For comparison, we used DAS with a single SOS to reconstruct the image. We selected two regions of interest (ROIs), indicated in orange and blue boxes, and adjusted the SOS (1550 m/s and 1565 m/s) to achieve the best imaging effect for the microspheres in the two regions, respectively. From the imaging results, it can be observed that traditional DAS method using a uniform and fixed SOS are difficult to achieve overall focusing effects, while the TI-MDAS algorithm achieves simultaneous focusing effects on both ROIs. Additionally, as can be found in the intensity profile shown in Fig. 12(b), the DAS result exhibits defocusing artefacts, which is well overcome by the TI-MDAS method.

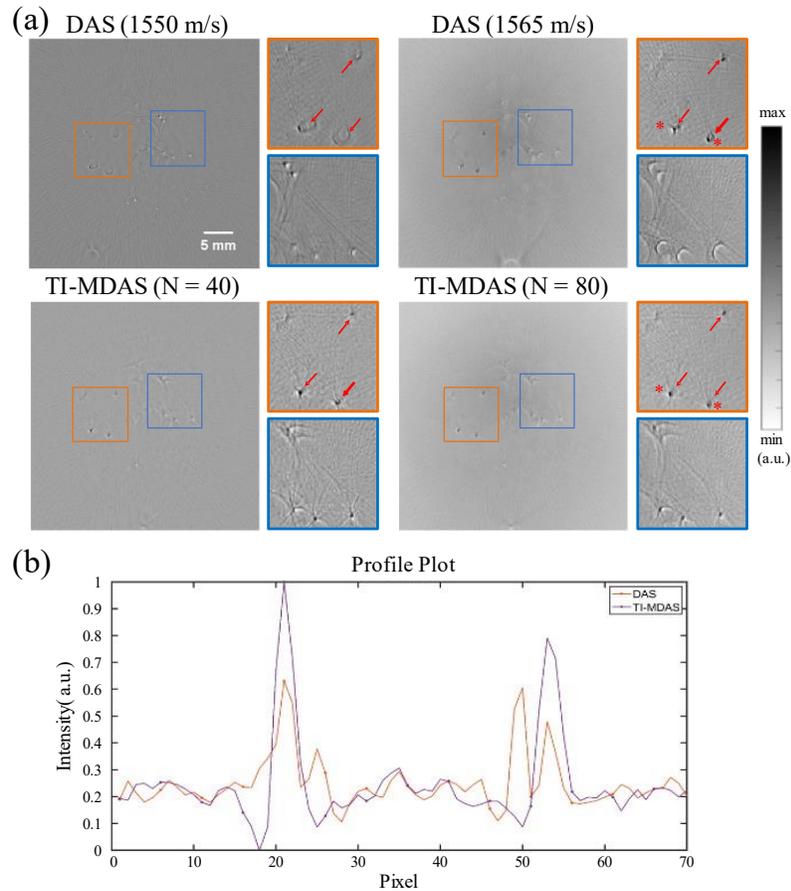

**Fig. 12.** Phantom experiment. (a) PAT image reconstruction results using DAS algorithm (1550 m/s, 1650 m/s) and our TI-MDAS algorithm. (b) The PAT image profiles along the line connecting the two red stars indicated in the orange region in (a).

### C. Ex vivo tissue imaging results

The reconstructed results of the SOS distribution in excised biological tissues are shown in Fig. 13. As illustrated, when N is 40, the reconstructed SOS distribution can already identify large areas of muscle tissue (black arrow) and small areas of adipose tissue (red arrow). When N is 80, the reconstructed SOS distribution can clearly distinguish the adipose tissue regions, and the edges of the muscle tissue regions are more distinct, showing a clear transition in SOS, which is also consistent with the actual situation.

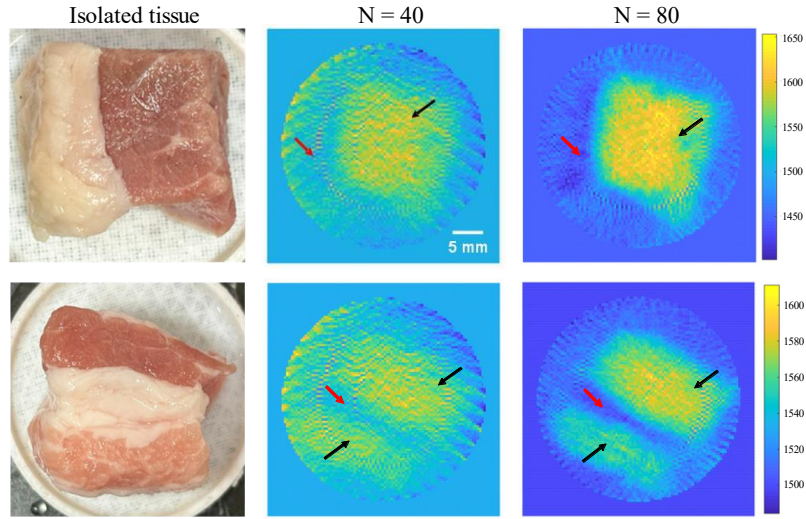

**Fig. 13.** Ex vivo tissue imaging experiment. Reconstruction results of SOS distribution of lamb and pork tissue under different numbers of scanned PRB positions.

The PAT image reconstruction results of excised biological tissues are shown in Fig. 14(b). Similarly, we compare the traditional single-SOS DAS reconstruction results (with the best visual effect achieved when SOS is 1515 m/s) with the TI-MDAS reconstruction results. It can be observed that the overall PAT reconstruction quality of TI-MDAS is better. Furthermore, the peaks in the image intensity profile (Fig. 14(c)) indicate that the reconstructed image of TI-MDAS has better focusing effects.

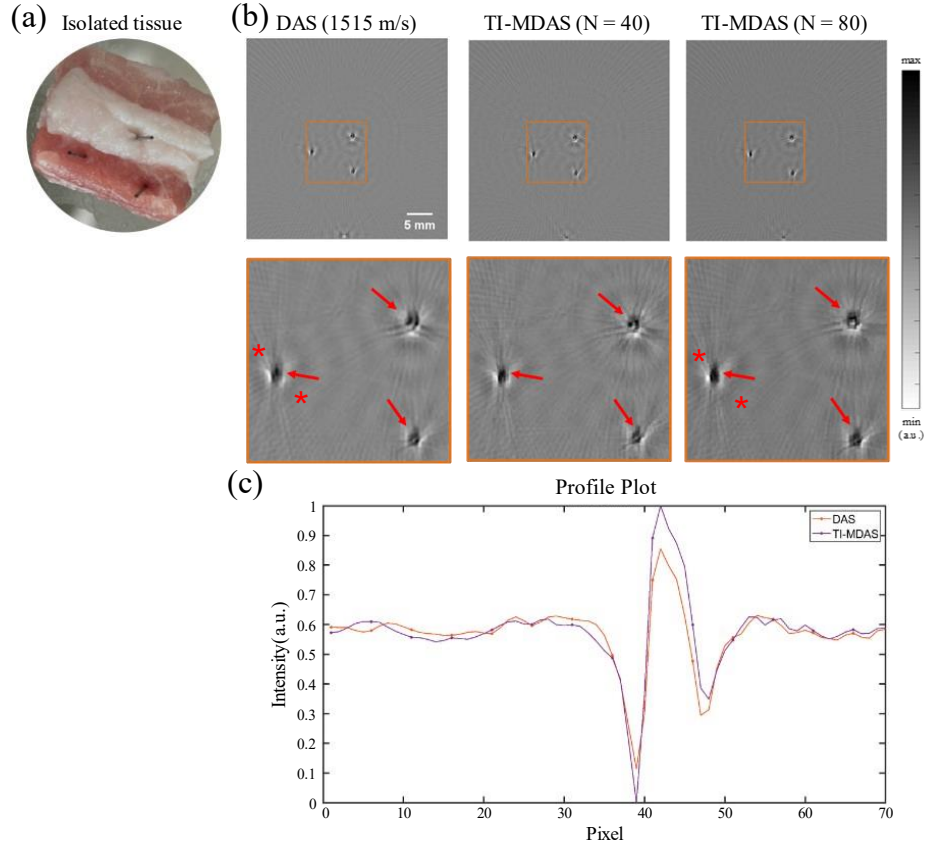

**Fig. 14.** (a) Photograph of the ex vivo pork tissue phantom with three carbon rod insertions. (b) PAT image reconstruction results of the tissue sample using DAS algorithm (1515 m/s) and TI-MDAS algorithm. (c) Image profiles along the line connecting the two red stars indicated in (b).

### D. Computational efficiency of TI-MDAS algorithm

In our experiments, we used MATLAB R2019a as the testing platform, running on a computer equipped with a 4-core processor at 2.20 GHz and 32 GB of RAM. We selected a coarse grid resolution of 100×100 and a fine grid resolution of 400×400. The average time taken to calculate TOF on the fine grid was around 64 seconds, whereas the average time for a single calculation of TOF using TI-MDAS interpolation was 1.6 seconds, indicating that TI-MDAS only required 1/40th of the time compared to direct calculation. When randomly selecting a target pixel, the average TOF on the fine grid was $1.7532 \times 10^{-2}$ ms, while the TOF calculated through TI-MDAS was $1.7543 \times 10^{-2}$ ms. The average error between the two methods for TOF calculation was $1.1 \times 10^{-5}$ ms, which is significantly smaller than three orders of magnitude of the TOF itself. Therefore, the TI-MDAS algorithm achieves a significant computational acceleration with minimal error.

## 5. Discussion

The reconstruction and imaging of heterogeneous SOS can aid in clinical diagnosis by measuring variations in SOS within biological tissues [16, 17]. Therefore, by monitoring and imaging the SOS distribution within biological tissues, we can gain insights into disease progression, assisting diagnosis and prompting initiation of therapeutic measures. In this work, we present a method for reconstructing the cross-sectional SOS distribution based on PAT. Compared to conventional SOS reconstruction method based on ultrasound tomography, our approach solely relies on a PAT system to acquire the SOS distribution. By adding the SOS imaging capability, this work enriches the existing PAT imaging modality.

Our PRB method relies on small photoacoustic absorbing target that served as SOS imaging beacon. By scanning the beacon around the object, our method probes the internal SOS structure non-invasively. We discretize the SOS of the imaged object into pixels and devise a model-based inversion method for SOS reconstruction. Leveraging the Nesterov's accelerated gradient descent algorithm, we iteratively solve for the SOS distribution, achieving pixel-level reconstruction of the SOS.

Acoustic heterogeneity-induced imaging artefact has long been an unsolved problem in PAT imaging. Based on our PRB method, our work shows that accurate SOS distributions can significantly enhance PAT image reconstruction quality and mitigate reconstruction artifacts. The proposed TI-MDAS algorithm utilizes the obtained SOS distribution to more accurately calculate signal propagation time, enabling more precise localization of absorbers. By reducing image distortions and artifacts while enhancing image quality, this, in turn, further validates the effectiveness of our PRB method.

While the PRB approach achieves pixel-level tomographic SOS reconstruction, it relies on multiple scanning of the PRB, and thus increases the SOS and PAT imaging time period. Additionally, our current method is based on ring-shaped array detector. The feasibility of other detection geometries, such as those with hemisphere detector[40, 41], remains to be further validated.

## 6. Conclusions

In this work, we propose a cross-sectional SOS imaging method based on photoacoustic reversal beacon, or PRB. Our method does not rely on ultrasonic imaging system but solely utilizes PAT to achieve pixel-level SOS reconstruction non-invasively. Furthermore, based on the obtained SOS map, we have implemented a TI-MDAS algorithm for acoustic corrected PAT image reconstruction. The effectiveness and reliability of our method are validated through simulation, phantom, and biological tissue imaging experiments.

**Acknowledgements**

This work was supported in part by National Natural Science Foundation of China (62371220), Guangdong Basic and Applied Basic Research Foundation (2021A1515012542, 2022A1515011748).